\documentclass{emulateapj}
\usepackage{latexsym}
\usepackage{dcolumn}
\usepackage{bm}
\input{epsf}
\newcommand{\be}{\begin{equation}}
\newcommand{\ee}{\end{equation}}
\newcommand{\ba}{\begin{eqnarray}}
\newcommand{\ea}{\end{eqnarray}}
\newcommand{\bi}{\begin{itemize}}
\newcommand{\ei}{\end{itemize}}
\newcommand{\bfi}{\begin{figure}
\epsfxsize=9cm
\epsffile}
\newcommand{\efi}{\end{figure}}
\newcommand{\mnras}{MNRAS}
\newcommand{\araa}{ARAA}
\newcommand{\apjl}{\apj}
\newcommand{\apjs}{\apj Suppl.}

\begin{document}
\title{The probability distribution function of the
  SZ power spectrum: an analytical approach}
\author{Pengjie Zhang}
\email{pjzhang@shao.ac.cn}
\affil{Shanghai Astronomical Observatory, Chinese Academy of
  Science, 80 Nandan Road, Shanghai, China, 200030}
\affil{Joint Institute for Galaxy and Cosmology (JOINGC) of
SHAO and USTC}
\author{Ravi K. Sheth}
\affil{209 S. 33rd Street, Department of Physics and Astronomy,       
University of Pennsylvania, Philadelphia, PA 19104, USA}

\begin{abstract}
The Sunyaev Zel'dovich (SZ) signal is highly non-Gaussian, so the 
SZ power spectrum (along with the mean $y$ parameter) does not 
provide a complete description of the SZ effect.  Therefore, SZ-based 
constraints on cosmological parameters and on cluster gastrophysics 
which assume Gaussianity will be biased.  
 We derive an analytic expression for the $n$-point joint PDF of 
the SZ power spectrum.  Our derivation, which is based on the halo 
model, has several advantages: 
 it is expressed in an integral form which allows quick computation; 
 it is applicable to any given survey and any given angular scale; 
 it is straightforward to incorporate many of the complexities 
 which arise when modeling the SZ signal.  
To illustrate, we use our expression to estimate $p(C_\ell)$, 
the one-point PDF of the SZ power spectrum. 
For small sky coverage (applicable to BIMA/CBI and the Sunyaev 
Zel'dovich Array experiments), our analysis shows that $p(C_\ell)$ 
on the several arc-minute scale is expected to be strongly skewed, 
peaking at a value well below the mean and with a long tail which 
extends to tail high $C_\ell$ values.  In the limit of large sky 
coverage (applicable to the South Pole Telescope and Planck), 
$p(C_\ell)$ approaches a Gaussian form.  However, even in this 
limit, the variance of the power spectrum is very different from 
the naive Gaussian-based estimate.  This is because different 
$\ell$ models are strongly correlated, making the cosmic variance 
of the SZ band-power much larger than the naive estimate.  
Our analysis should also be useful for modeling the PDF of the 
power spectrum induced by gravitational lensing at large $\ell$.  
\end{abstract}

\section{Introduction}
The Sunyaev Zel'dovich (SZ) effect \citep{Zeldovich69,Sunyaev72}
is a powerful probe of baryons, dark matter and dark energy 
\citep{Birkinshaw99,Carlstrom02}.  To extract the full statistical 
power of this probe, SZ statistics must be understood to an 
accuracy which matches that which the observations will reach.  
The statistic which has been most actively studied is the ensemble average power 
spectrum $\bar C_\ell$ of the thermal SZ signal.  
For the simplest case of adiabatic gas evolution, both analytic 
\citep{Cole88,Makino93,Atrio-Barandela99,Komatsu99,Molnar00,Majumdar01,Zhang01,Komatsu02}
and numerical 
\citep{daSilva00,Refregier00,Seljak01,Springel01,Zhang02} 
approaches are beginning to give consistent results 
\citep{Refregier02,Komatsu02}.

However, $\bar{C}_\ell$ alone (or in combination with the mean $y$ 
parameter) does not provide a complete description of the SZ effect. 
This is because the SZ signal is strongly non-Gaussian 
\citep{Cooray01,Seljak01,Yoshida01,Zhang02,Komatsu02}. 
In contrast to the primary CMB and weak gravitational lensing signals, 
the SZ signal is non-Gaussian even on large scales \citep{Zhang01}.  
This is because the effect is dominated by rare massive clusters.  
Therefore, interpretation of a SZ $C_\ell$ measurement requires not 
only a robust prediction of the expected ensemble-averaged 
$\bar{C}_\ell$, but also the probability distribution function (PDF) 
of $C_\ell$ around this mean value.

Because the SZ effect is dominated by the most massive clusters, and 
the abundance of such clusters is very sensitive to the amplitude of 
density fluctuations \citep{Press74,ST99}, the SZ power spectrum depends 
strongly on $\sigma_8$:  $C_l\propto \sigma_8^{\sim 7}$ for the range 
of currently interesting cosmological models 
\citep{Seljak01,Zhang01,Komatsu02}.  In principle, this makes 
the SZ power spectrum a powerful probe of $\sigma_8$.  
The CMB power spectra excess reported by CBI
\citep{Mason03,Readhead04} and BIMA   
\citep{Dawson02} are consistent with having a contribution from the 
SZ effect provided $\sigma_8\simeq 1.0$
\citep{Dawson02,Bond05,Readhead04}.  This SZ 
explanation is further supported by the frequency dependence of the CMB power
excess \citep{Kuo06}. 
However, this value is uncomfortably high compared to the WMAP3 
value of $\sigma_8\approx 0.8$ \citep{Spergel06}.

The SZ-based estimates of $\sigma_8$ assume adiabatic evolution of 
the gas.  Feedback \citep{daSilva01,White02,Lin04} and 
cooling \citep{daSilva01,Zhang03} effects can decrease the SZ 
power spectrum by a factor of 2; if these are included, then the 
required value of $\sigma_8$ increases further.   
Other mechanisms such as magnetic fields \citep{Zhang04} can further 
reduce the amplitude of $C_l$.  Other contributions to the SZ signal 
have also been studied---the SZ effect of the first stars \citep{Oh03}, 
and contributions from unvirialized intergalactic medium 
\citep{Atrio-Barandela06}.  But if the WMAP 3 estimates are accurate, 
$\tau\simeq 0.1$ and $\sigma_8\simeq 0.8$, then these additional 
contributions are likely to be subdominant.  
Thus to explain the CBI/BIMA SZ measurements it appears that 
$\sigma_8\ga 1.1$-$1.2$:  evidently, a severe discrepancy with 
the CMB-based estimate exists \footnote{Contaminations of unresolved
  and unremoved point sources remain a possible solution to the
  power excess problem \citep{White04,Toffolatti05,Douspis06},
  although current modeling is quite uncertain.}.

There are at least two ways in which the discrepancy can be 
decreased.  One is to assume that the uncertainty on the estimate 
is artificially small.  This would happen, for instance, if the 
statistical errors on the SZ $C_\ell$ and  $C_{\ell^{'}}$ are 
highly correlated, even for widely separated $\ell,\ell^{'}$ pairs.  
If so, then the cosmic variance of the measured band-power SZ 
$C_\ell$ would be much larger than the naive Gaussian-based 
estimates.

Alternatively, the SZ-based estimates of $\sigma_8$ 
arise from requiring the measured $C_\ell$ to match the ensemble 
averaged power spectrum predicted from theory.  However, if the 
distribution of $C_\ell$ is skewed, then this requirement is 
unreasonable.  Since we have good reason to expect the signal to be 
non-Gaussian, it may be that the discrepancy can be resolved if one 
accounts for this non-Gaussianity.  For small sky coverage in particular, 
$p(C_\ell)$ can be highly skewed, with a non-negligible tail toward 
high $C_l$.  Thus, for a survey with small sky coverage, the probability 
of obtaining $C_\ell$ several times larger the mean $\bar C_\ell$ 
is not negligible.  
\citet{Dawson06} simulated an SZ map for one choice of cosmological
model and then scaled the simulated SZ decrement (assuming the 
signal scales as $\sigma_8^{7/2}$, so that $C_\ell$ scales as 
$\sigma_8^7$) to estimate the effect of this non-Gaussianity 
as $\sigma_8$ varies.  They used this procedure to argue that 
$\sigma_8\sim 0.7$ may well produce a long enough tail of $C_\ell$ 
values to explain the CBI/BIMA measurement.  While this is 
reassuring, it is not obvious that this simple rescaling is indeed 
an accurate description of how the non-Gaussian distribution of 
$C_\ell$ depends on $\sigma_8$.

The discussion above shows why understanding the full PDF of $C_\ell$ 
is crucial. Current theoretical understanding of $p(C_\ell)$ is 
limited.  Even for a single cosmological model and a given sky 
coverage (e.g. $1$ deg$^2$ or larger), simulations lack sufficient 
realizations to measure the full PDF reliably.  
The scaling method of \citet{Dawson06} only provides a crude 
estimate, since the non-Gaussianity (and the PDF) is certainly 
a function of $\sigma_8$. Analytical estimates of the low order 
moments of $p(C_\ell)$, such as the variance and covariance 
of $C_\ell$ have been made \citep{Cooray01,Zhang01,Komatsu02}. 
These calculations provide quantitative estimates of the errors 
in $C_\ell$, and represent important first steps towards understanding 
the SZ non-Gaussianity. However, to specify $p(C_\ell)$, higher order 
moments are also required.  The computational time required to 
calculate higher order moments rapidly becomes prohibitive. 
To estimate $p(C_\ell)$ for any given cosmology and given survey 
strategy in a reasonable amount of time, going beyond this moment 
by moment calculation is essential.

This paper presents a fast, complete, analytical method for 
calculating the full $n$-point PDF. It is based on the halo model 
\citep{Cooray02} and allows one to easily estimate the effect on 
the SZ signal of different treatments of the relevant gas physics. 
\S \ref{sec:variance} outlines the halo model calculation of the 
lowest order statistics of the SZ power spectrum: 
the mean, the variance and the covariance.  
An analytical expression for the $n$-point joint PDF of the 
SZ power spectrum is derived in \S \ref{sec:pdf}.  We show 
explicitly that it correctly reproduces the usual expressions for 
the mean, variance and covariance.  We then show that the analysis 
is particularly simple in the limits of very large and very small 
sky coverage. 
A numerical calculation of the one-point PDF, and a comparison 
with the large sky coverage limit is presented in \S \ref{sec:numerical}.  
\S \ref{sec:summary} summarizes our results, highlights several 
key simplifications in our approach, and discusses possible extensions.

Where necessary in this paper, we adopt a flat $\Lambda$CDM cosmology 
with $\Omega_m=0.3$, $\Omega_{ \Lambda}=0.7$, $\sigma_8=0.9$, and 
$H_0 = 100h$~km~s$^{-1}$~Mpc$^{-1}$ with $h=0.7$.   
To illustrate our results we use an initial power spectrum with 
index $n=1$, the BBKS fitting formula for the transfer function, 
the NFW profile \citep{Navarro96} for the dark matter profile, 
the Press-Schechter formula \citep{Press74} for the halo mass 
function and the associated halo bias factor \citep{Mo96} when 
modeling halo clustering.  
When our formalism is used to interpret simulations or observations, 
more accurate models of halo abundance and clustering, such as those 
of \citet{ST99}, should be used.

\section{The mean, variance and covariance of the SZ power spectrum}
\label{sec:variance}
The dominant contribution to the SZ power spectrum is from virialized 
regions \citep{Seljak01,White02,Cooray02,Carlos06}. This makes the 
halo model particularly well suited for estimating $C_\ell$ and its 
distribution. 
On the angular scales of interest in what follows, the two-halo term 
is always much smaller than the one-halo  term:  the contribution is 
less than $1\%$ at $\ell\ga 1000$ \citep{Komatsu99}, so it can be 
neglected.  Thus, the measured power spectrum 
$C\equiv C({\bf \ell})|_S$ at multipole ${\bf \ell}$, is just a 
sum over the contributions from each cluster in the survey.  
I.e., for a survey with sky coverage $f_{\rm sky}$, 
\be
\label{eqn:C}
 C|_S = f_{\rm sky}^{-1}\int dz \frac{dV}{dz} 
  \int dM\frac{d{n}}{dM}  S(M,z;\lambda)\rho(\lambda|M,z)d\lambda\ .
\ee
Here, $V$ is the survey volume. The explicit factor $f_{\rm sky}^{-1}$ in the
  above equation  cancels
  the implicit dependence of $V$ on $f_{\rm sky}$  ($V\propto f_{\rm sky}$) so the
  expectation value of $C|_S$ does not depend on $f_{\rm sky}$. $S(M,z;\lambda)$ is the SZ power 
spectrum of a single cluster at the corresponding ${\bf \ell}$. 
In this and the following expressions, we assume that no confusion 
will arise from the fact that we do not write explicitly that 
both $S$ and $C$ depend on ${\bf \ell}$.   
The quantity $S$ describes the structure of the cluster; it is 
mainly determined by the cluster mass $M$ and redshift $z$, but it 
also depends on other parameters $\lambda$, which may include the 
cluster concentration parameter $c$, cluster shape, etc. 
These parameters may or may not correlate tightly with $M$, and, in
general, they almost certainly are not fixed by $M$ alone. 
We use $\rho(\lambda|M,z)$ to denote the distribution of these 
parameters at fixed $M$ and $z$; it is normalized so that 
$\int \rho(\lambda|M,z)\,{\rm d}\lambda = 1$.  
For simplicity, we neglect these extra parameters in the numerical 
calculations which follow, although we do include them in our 
analytical expressions.  We will discuss their possible effects 
later.

If $d\bar{n}/dM$ is the ensemble averaged halo mass function, 
then the comoving number density of clusters per mass interval 
that are actually in the survey volume is 
\begin{equation}
 dn/dM = d\bar{n}/dM\,\Bigl[1+\delta_L(M)+\delta_P\Bigr] \ .
\end{equation}  
Here, $\delta_L(M)$ denotes the mean fluctuation in cluster 
number density which arises from the fact that the density 
of dark matter in the survey volume may not be exactly the same 
as the universal average, and $\delta_P$ represents the fluctuation 
which arises from the fact that clusters represent a discrete 
point process realization of the smooth density field 
$d\bar{n}/dM\,[1+\delta_L(M)]$.  It is standard to assume that 
$\delta_P$ is drawn from a Poisson distribution with mean 
$d\bar{n}/dM\,[1+\delta_L(M)]$ \citep{Casas02,Hu06,Holder06}.

In what follows, we will often consider bins in redshift, mass 
and $\lambda$.  We use the Latin letter $j$ to denote $j$th 
redshift bin and the Greek letter $\alpha$ to denote the 
$\alpha$th bin of $M$ and $\lambda$. 
The expectation value of $C|_S$ is 
\be
\label{eqn:meanC}
 \bar{C}=\int dz \frac{dV^f}{dz} \int dM \frac{d\bar{n}}{dM}\,
              S(M,z;\lambda)\,\rho(\lambda|M,z)\,d\lambda,
\ee
where $V^f$ is the comoving volume of the survey.  
The variance  of $C|_S$ is the sum of two terms:  
\ba
\label{eqn:variance}
 \sigma_C^2 &=& f^{-1}_{\rm sky}\int dz \frac{dV^f}{dz}
    \int dM \frac{d\bar{n}}{dM}\, S^2(M,z;\lambda)\,\rho(\lambda|M)d\lambda
    \nonumber\\
    && \quad +\sum_j \sigma_j^2 \left[\int dM \frac{d\bar{n}}{dM}\,
                    b(M,z_j) S(M,z_j;\lambda)\right. \nonumber \\
&&\qquad\qquad\qquad\quad\times\quad 
  \rho(\lambda|M,z_j)\,d\lambda \ \Delta V^f_j\Bigr]^2 \ .
\ea
The first term on the right hand side is often called the Poisson 
term. If 
\be
 N_{\rm P}\equiv \left[\int S^2\rho\, d\bar{n}\, dV^f\, d\lambda\right]^{-1}
                 \left[\int S \rho\, d\bar{n}\, dV^f\, d\lambda\right]^2
 \label{defNp}
\ee
denotes an effective number of clusters, then the fractional 
fluctuation in $C_\ell$ induced by the Poisson term is 
 $[N_{\rm P}f_{\rm sky}]^{-1/2}$.

The second term on the right hand side of equation~(\ref{eqn:variance}) 
accounts for cluster clustering.  The sum is over all redshift bins $j$. 
We choose bin sizes $\Delta z\sim 0.1$ so that $\delta_{\rm S}$, 
the mass overdensity fluctuation smoothed over the survey volume 
$f_{\rm sky}\Delta V^f$ in any given redshift bin, is (approximately) 
uncorrelated with $\delta_{\rm S}$ in other bins.  The quantity 
\begin{equation}
 \sigma^2_j\equiv \left\langle\delta_{\rm S_j}^2\right\rangle 
              = \int P_j(k)\,W_j^2({\bf k})\,d^3k/(2\pi)^3
\end{equation}
is the rms of the matter over-density when smoothed over the survey 
window in the $j$th redshift bin; $P_j$ and $W_j$ represent 
the 3D matter density power spectrum, and the Fourier transform 
of the survey window function, both in the $j$th redshift bin.  
The factor $b$ represents the fact that the cluster distribution is 
biased relative to the dark matter:  $\delta_L(M) = b(M)\delta_{\rm S}$.  
\bfi{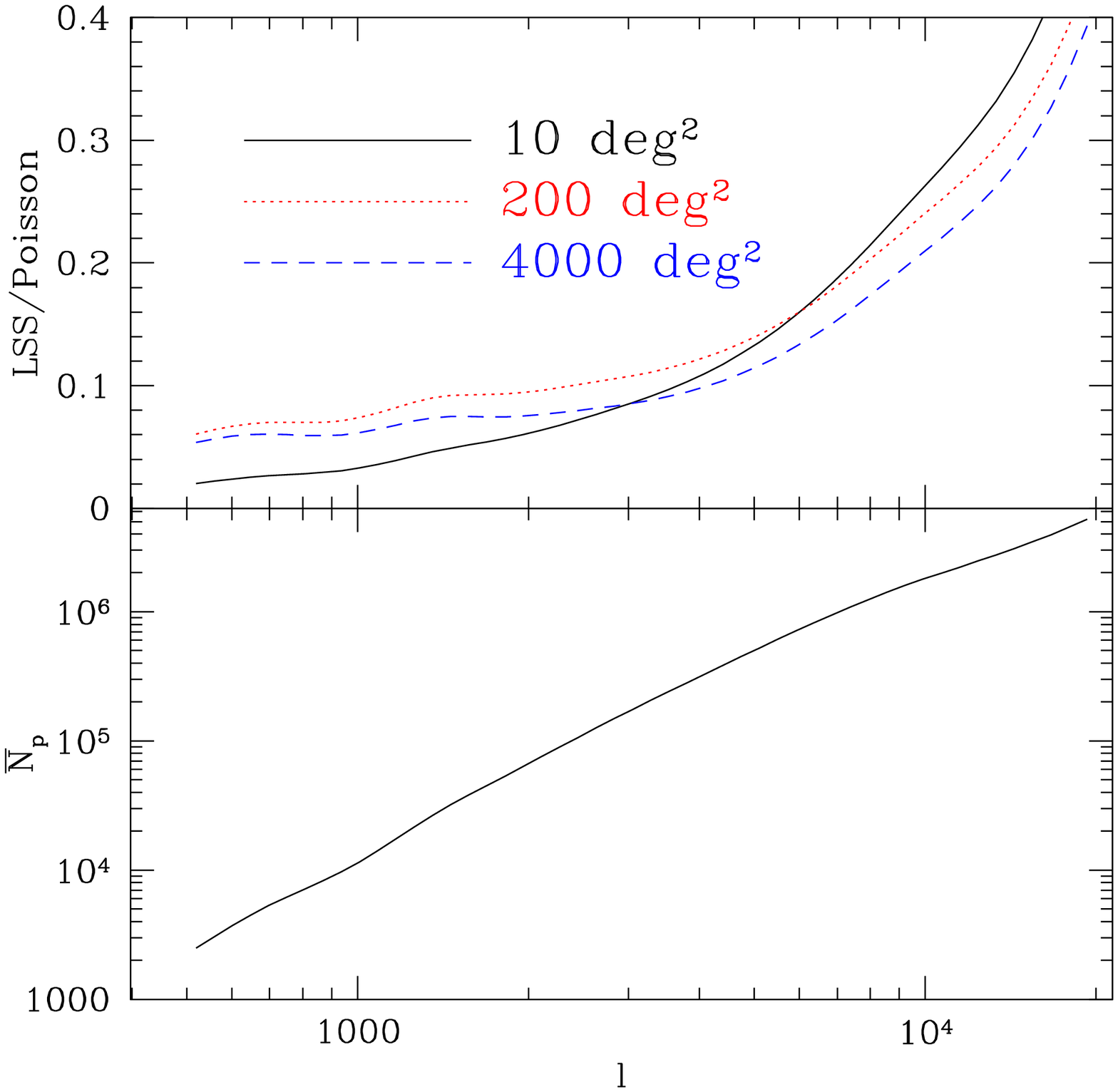}
 \caption{The variance of the SZ power spectrum. 
  The lower panel shows the weighted cluster number $\bar{N}_P$. 
  $(N_{\rm P}\,f_{\rm sky})^{-1/2}$ is the fractional statistical 
  error caused by the  Poisson distribution of clusters. 
  The upper panel shows the ratio between the statistical error 
  caused by the large scale clustering and the Poisson distribution. 
  We show below that when $N_{\rm P}\,f_{\rm sky}\gg 10$, then $p(C_\ell)$ 
  will approach a Gaussian form, in accordance with the central limit 
  theorem. \label{fig:variance}}
\efi

We model $S(M,z)$ using the universal gas profile of \citep{Komatsu01}. 
This model makes three reasonable approximations/assumptions:
 (1) Intra-cluster gas is in hydrostatic equilibrium; 
 (2) the gas pressure $P_g\propto \rho_g^{\gamma}$ and the 
     polytropic index $\gamma$ is a constant; 
 (3) the ratio of the gas density $\rho_g$ and the dark matter 
     density approaches the cosmological value $\Omega_b/\Omega_{\rm DM}$ 
     in the outer regions of clusters. 
Given a dark matter profile, this model predicts $S$ uniquely. 
When combined with a model for the mass function $d\bar{n}/dM$, 
this model completely specifies $\bar{C}$ \citep{Komatsu02}.

Fig.~\ref{fig:variance} shows the Poisson and halo clustering 
contributions to the SZ power spectrum variance for a range 
of choices of the fraction of sky covered $f_{\rm sky}$. 
The top panel shows that the halo clustering contribution is 
sub-dominant, but non-negligible, at the relevant scales.  
Although the Poisson term has a simple dependence on the sky 
coverage ($\propto f^{-1}_{\rm sky}$), the halo clustering term 
is more complicated.  It depends implicitly on $f_{\rm sky}$ because 
$\sigma^2$ depends on the shape of the survey volume.  
Since  the effective power index of CDM power spectrum $n_{\rm eff}$ 
varies with scale, $\sigma^2$ has non-trivial dependence on the
smoothing scale. Thus the ratio between the halo clustering term 
and the Poisson term has a complicated dependence on $f_{\rm sky}$.

The bottom panel shows that $N_{\rm P}$, the effective number of clusters, 
increases with increasing $\ell$.  This reflects the fact that the 
small scale SZ power spectrum is dominated by smaller halos
(e.g. \cite{Komatsu02}). Since the Poisson contribution to the 
power scales as $N_{\rm P}^{-1/2}$, it decreases as $\ell$ increases.
Because the bias factor also decreases with decreasing halo mass, 
the halo clustering contribution also decreases on small scales.  
However, the decrease as not as dramatic as for the Poisson term; 
this is why the ratio of this to the Poisson term increases with 
$\ell$.

The {\it universal} gas model is spherically symmetric, so different 
${\bf \ell}$ with the same $\ell\equiv |{\bf \ell}|$ are 
identical.\footnote{Accounting for cluster ellipticities breaks 
this degeneracy as we discuss in \S \ref{sec:discussion}.}  
Hence, the combination of $(2\ell+1)$ modes all having the same 
$|{\bf \ell}|$ has the same statistical error $\sigma_C$ and the
fractional error $\simeq [N_{\rm P} f_{\rm sky}]^{-1/2}$. 
For a Gaussian random field, the {\it fractional} statistical error 
of the power spectrum for each $\ell$ is
 $[(2\ell+1)f_{\rm sky}/2]^{-1/2}$. 
Since $N_{\rm P}\gg \ell$ (Fig. \ref{fig:variance}), the statistical error 
of the SZ effect of each $\ell$ is {\em smaller} than it would be 
in a Gaussian random field.  The halo model shows that this is 
because clusters are nearly self similar, so the statistical error 
in the SZ power spectrum is mainly determined by the fluctuations in 
cluster number. Because there are many clusters across the sky, these 
fluctuations are small, and the resulting statistical error is smaller 
than the corresponding Gaussian estimate. 
\bfi{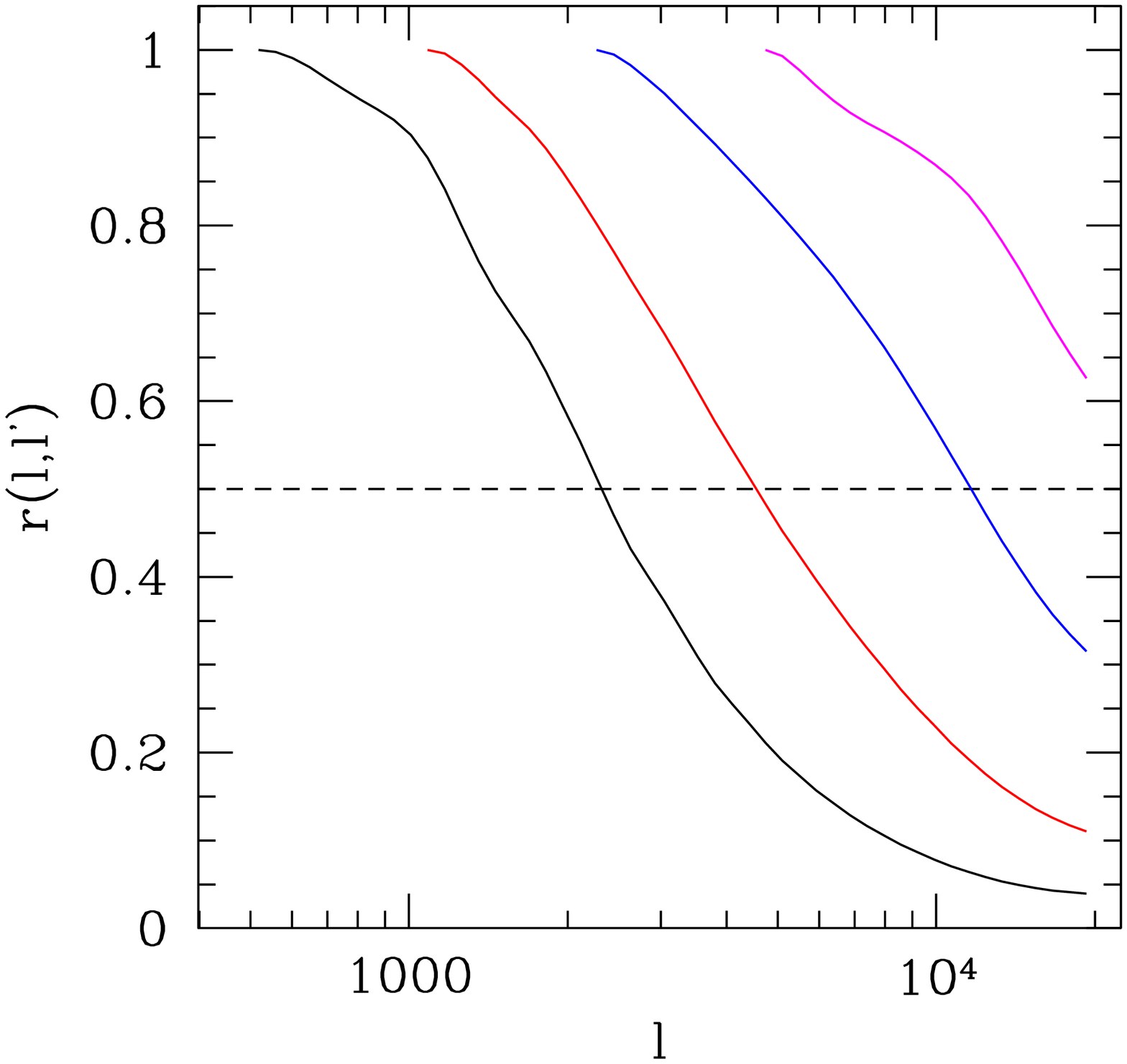}
\caption{The cross correlation coefficient $r$ between different 
  $\ell$ modes.  The dependence of $r$ on sky coverage is 
  negligible; we have assumed coverage of $200$ deg$^2$. 
  The value of $\ell^{'}$ for each line is indicated by the 
  lowest $\ell$ value for which it is plotted. \label{fig:r} }
\efi

However, this does {\em not} mean that the statistical precision 
of the SZ measurement exceeds the Gaussian limit.  Although 
clusters are nearly self similar, the contribution of each cluster 
spans a large range in $\ell$.  As a result, the observed signal 
in the different $l$ modes can be highly correlated.  If $\Delta \ell_c$
denotes the correlation length in $\ell$ space,  
then it is the combined error on $\Delta \ell_c$ modes which is 
the same as that of one single mode.  This can be much larger 
than the corresponding Gaussian error 
$[(2\ell+1)\,\Delta \ell_c\, f_{\rm sky}/2]^{-1/2}$.

To estimate $\Delta\ell_c$ we compute the covariance between 
$C_\ell$ and $C_{\ell^{'}}$:  
\ba
\label{eqn:covariance}
 {\rm Cov}_{\ell\ell^{'}} &=& f^{-1}_{\rm sky}
  \int dz \frac{dV^f}{dz}\, \int dM\, \frac{d\bar{n}}{dM}\,
    SS^{'}\rho(\lambda|M)d\lambda \nonumber\\
 &+& \sum_j \sigma_j^2 \left[\int dM \frac{d\bar{n}}{dM}\,
       b\,S\, \rho(\lambda)\,d\lambda\, \Delta V^f_j\right]\nonumber\\
 &&\qquad\times\quad \left[\int dM \frac{dn}{dM}\, b\,S^{'}\, 
             \rho(\lambda)\,d\lambda \,\Delta V^f_j\right]
\ea
where $S^{'}\equiv S_{\ell^{'}}(M,z;\lambda)$.  
Again, the Poisson term dominates this covariance. 
The cross correlation coefficient between the statistical errors 
of $C_\ell$ and $C_\ell'$ is
\be
\label{eqn:r}
 r^2\equiv \frac{{\rm Cov}_{\ell\ell^{'}}}{\sigma_C^2\sigma_{C^{'}}^2}\ .
\ee
Since the Poisson term dominates in both the variance and covariance 
term, the dependence of $r$ on $f_{\rm sky}$ is negligible.

Fig. \ref{fig:r} shows $r$ for $200$ deg$^2$ sky coverage. 
Note that the correlation length $\Delta\ell_c\ga 1000$. In
simulations, the finite size of the SZ map $\theta_{\rm map}$, 
makes the natural bin size $\Delta\ell=2\pi/\theta_{\rm map}\sim 300$ 
for typical $\theta_{\rm map}\sim 1^{\circ}$. Thus, $\ell$ modes 
over this $\Delta\ell$ can be treated as completely correlated. 
The fractional statistical error on $C$ measured in simulations, 
with respect to the Gaussian random field, is expected to be 
 $\sim [\ell\Delta\ell/N_{\rm P}]^{1/2}$. 
This is indeed what is seen.  For example, the analysis above 
suggests that the quantity shown in Fig.6 of \citet{Zhang02} is 
 $a_4\simeq (2\ell+1)\Delta\ell [1+{\rm LSS}/{\rm Poisson}]/N_{\rm P}-2$ 
(if the contribution of halo clustering has been taken into account).  
The numbers from Fig. \ref{fig:variance} suggest that $a_4\sim 50$ 
at $\ell=10^3$ and $a_4\sim 2$ at $\ell=10^4$, consistent with the 
simulations \citet{Zhang02}.

\section{The PDF of the SZ power spectrum}
\label{sec:pdf}
The quantities $\bar{C}$, $\sigma_C$ and ${\rm Cov}_{\ell\ell^{'}}$ 
are the lowest order moments of the power spectrum PDF $p(C_\ell)$. 
However, because the SZ signal is non-Gaussian, these low order 
moments do not uniquely determine $p(C_\ell)$.  Since $p(C_\ell)$ 
is required to make unbiased error analyses and parameter constraints, 
this section provides an analytical integral formula for $p(C_\ell)$. 
This integral form, which is based on the halo model, allows fast 
numerical calculation and has the flexibility to incorporate much 
of the relevant physics. 
For clarity, \S~\ref{subsec:onezbin} considers the case of a single 
redshift bin; the contributions from different bins are summed 
in \S~\ref{subsec:projection}.

\subsection{Single redshift bin}\label{subsec:onezbin}
Given a survey and a redshift bin, the survey volume is fixed. 
We sort the clusters in this volume into bins of $M$ and, 
in principle, bins in the other parameters $\lambda_l$. 
Then the overall SZ power spectrum is 
\be
C=f^{-1}_{\rm sky} \sum_{\alpha} N_{\alpha}\, S_{\alpha}\ ,
\ee
where $N_{\alpha}$ is the number of clusters in the $\alpha$-th bin 
and $S_{\alpha}$ is the SZ power spectrum of each cluster in the 
$\alpha$-th mass (and $\lambda$) bin.  
The PDF of $C$ in the $j$th redshift bin is 
\ba
 p_j(C) &=& \sum_{N_1=0}^{\infty}\cdots
          \sum_{N_{\alpha}=0}^{\infty}P(N_1,\ldots,N_{\alpha})\\
      &&\qquad\times \quad \delta_D\left(f^{-1}_{\rm sky}\, 
          \sum_{\alpha} N_{\alpha}\,S_{\alpha}-C\right)\ ,
\ea
where $P(N_1,\ldots,N_\alpha)$ is the joint distribution of the 
$N_\alpha$.

This distribution is computed in two steps.  
First, if the matter overdensity $\delta$ in the survey volume is 
not exactly zero, then the mean number of expected clusters is 
altered \citep{Mo96,Sheth02}.  We use 
\ba
 N_{{\alpha},L}=\bar{N}_{\alpha}(1 + \delta_{\alpha})
\ea
to denote this altered mean.  We then convert the smooth field 
$N_{{\alpha},L}$ to a discrete point process by assuming a 
Poisson sampling model.  Namely, we assume that given $\delta_\alpha$,  
\be
 P_{\rm P}(N_{\alpha}|\delta_{\alpha}) = 
  \frac{[N_{{\alpha},L}]^{N_{\alpha}}}{N_{\alpha}!}\exp[-N_{{\alpha},L}];
\ee
note that this Poisson process in the $\alpha$th bin is independent 
of that in the other bins.  Hence, 
\ba
 P(N_1,\cdots,N_{\alpha})&=&\int \left[\prod_
  {\alpha}P_{\rm P}(N_{\alpha}|\delta_{\alpha})\right]\\
 &&\quad\times \quad
   P_{\rm L}(\delta_1,\cdots,\delta_{\alpha})\prod_{\alpha} d\delta_{\alpha}\ .
   \nonumber
\ea
To proceed, we need a model for $P_{\rm L}(\delta_1,\cdots,\delta_{\alpha})$, 
the joint PDF of $\delta_{\alpha}$.

When the volume is small, stochastic discreteness effects dominate, and 
$P_{\rm L}(\delta_1,\cdots,\delta_{\alpha})\to
 \prod_{\alpha} P_{\rm L}(\delta_{\alpha})$. 
When this limit applies, the Poisson fluctuation is much larger than 
the matter fluctuation and the exact form of $P_{\rm L}$ is not necessary.  
However, $P_{\rm L}(\delta_1,\cdots,\delta_{\alpha})$ in the general case 
is more complicated \citep{SL99}.

When $\delta_\alpha\ll 1$, then the linear bias model 
\ba
 \delta_{\alpha} = b_{\alpha}\,\delta
\ea
works well \citep{Mo96,Sheth02}, and 
\ba
 P_{\rm L}\rightarrow P_{\delta}
  \prod_{\alpha}\delta_{\rm D}(\delta_{\alpha}-b_{\alpha}\delta)\ .
\ea
Typically, this requires the volume to be larger than about 
$(100h^{-1}$Mpc)$^3$.
In practice we will apply the linear bias model to all relevant cases.

To see what linear bias implies, note that the characteristic 
function is 
\ba
\label{eqn:biasP}
 \tilde{p}_j &\equiv& \int dC\,p_j(C)\exp({\rm i}fC)\\ \nonumber 
           &=&\int  P_{\rm L}(\delta_1,\cdots,\delta_{\alpha})\prod_{\alpha}\,
                    d\delta_{\alpha} \\
           &&\times \exp\left[\sum_{\alpha}
            N_{\alpha,{\rm L}}(e^{{\rm i}\tilde{f}S_{\alpha}}-1)\right], 
           \nonumber \\
           &=&\int d\delta\,P_j({\delta})\ 
         \exp\left\{\int \Bigl[\exp({\rm i}\tilde{f}
                    S(\lambda)-1\Bigr]dN_L\right\}\ \nonumber
\ea
where $\tilde{f}\equiv f/f_{\rm sky}$.  The second equality uses 
the Poisson model, and the final expression uses the linear bias 
model in the limit that the mass (and $\lambda$) bin size goes to 
zero.  I.e., $dN_{\rm L}=d\bar{N}(1+b\delta)$ is the expected number 
of clusters in the survey volume and redshift and mass bin (without 
Poisson fluctuation), and $d\bar{N}/f_{\rm sky}$ is the expected 
number of clusters averaged over the whole sky. 
Our notation makes explicit that the distribution $P_j(\delta)$ 
of the mass density fluctuations $\delta$ may depend on redshift.

On large scales, the distribution of $\delta$ should be approximately 
Gaussian:  $P_j(\delta)\propto \exp(-\delta^2/2\sigma_j^2)$. 
Hence, for each redshift bin,
\ba
\label{eqn:Pdelta}
\tilde{p}_j 
 &=&\exp\left(f_{\rm sky}D_j+\frac{f^2_{\rm sky}E_j^2\,\sigma_j^2}{2}\right)\ ,
\ea
where 
\be
 D_j = f^{-1}_{\rm sky} \int d\bar{N}\,(e^{{\rm i}\tilde{f}S}-1)\ ,\quad
\ee
and 
\be
 E_j = f^{-1}_{\rm sky} \int d\bar{N}\,b\, (e^{{\rm i}\tilde{f}S}-1)\ .
\ee 
Notice that neither $D_j$ nor $E_j$ have any dependence on 
$f_{\rm sky}$, because the implicit dependence of $\bar{N}$ on $f_{\rm sky}$
 ($\bar{N}\propto f_{\rm sky}$) cancels the $f_{\rm sky}^{-1}$ factor in the
 definitions of $D_j$ and $E_j$. 
We will soon show that $D_j$ describes the discreteness of the 
cluster distribution and $E_j$ describes the clustering effect.

Strictly speaking, equation~(\ref{eqn:Pdelta}) follows from integrating 
the Gaussian from $-\infty$ to $+\infty$.  Since $b\delta<-1$ is 
problematic (one may not have a negative number for the mean halo count 
in a cell), one might have thought the range of integration should be 
restricted to $\delta \ge -1/b$.  In practice, when the survey area is 
large, then $\sigma_j\ll 1$, so $\delta\ll 1$ and $b\delta>-1$ almost 
surely, and ignoring the restriction on $\delta$ is reasonable.  
This approximation simplifies the calculation in the next subsection 
significantly.

Before moving on, suppose that $S$ was the same constant for all 
clusters.  In this case, 
\ba
 f_{\rm sky}D_j &\to&\Bigl({\cal S}-1\Bigr)\int d\bar N 
                 \equiv \Bigl({\cal S}-1\Bigr)\,\bar N_{\rm eff} \nonumber\\
 f_{\rm sky}E_j &\to& \Bigl({\cal S}-1\Bigr)\int d\bar N\,b 
                 \equiv \Bigl({\cal S}-1\Bigr)\,\bar N_{\rm eff}\,b_{\rm eff},
\ea
where we have defined ${\cal S}\equiv e^{{\rm i}\tilde{f}S}$.  
Hence, 
\be
\label{eqn:unweighted}
 \tilde{p} \to
 \exp\left[({\cal S}-1)\,\bar N_{\rm eff} + 
  \frac{({\cal S}-1)^2}{2}\,\bar N_{\rm eff}^2\,b_{\rm eff}^2\sigma_j^2\right];
\ee
this expression is essentially the generating function of unweighted 
counts in cells.  Notice that it agrees with the expressions derived 
by \citet{Hu06} and \citet{Holder06} (although Hu \& Cohn set the 
lower limit of the integral over $\delta$ to $-1/b$).  
In this respect, our analysis generalizes their work to the case in 
which clusters contribute different weights. 

In the counts in cells case, suppose that each cluster is weighted by
the number of galaxies it contains, and that $W(S)$ is the generating
 function of galaxy counts per cluster.  Then the generating function of
 galaxy counts in cells, say $g(S)$, is given by
  $g(S) = G(W(S))$,
 where $G(S)$ is the generating function of unweighted cluster counts (e.g.
 our Eq. \ref{eqn:unweighted}).   We have used $W(S)$ to emphasize that one
 can think of the 
 galaxy counts in cells distribution as arising from a process where each
 cluster is weighted differently (i.e., by the number of galaxies in it).
 
 In our SZ calculation, each cluster has a different weight, so we can
 think of $W(S)$ as having a different value for each cluster.  So our
 Eq. \ref{eqn:biasP}  is indeed equivalent to the counts in cells  analogy.
 This analogy also shows how to generalize our formalism to include scatter in
 halo concentrations $\rightarrow$ scatter in SZ signal, $\rightarrow$ scatter
 in SZ $C_{\ell}$ even at  fixed halo mass.

One might have thought that equation~(\ref{eqn:Pdelta}) fails in the 
limit of small sky coverage.  Appendix \ref{sec:appendix} shows that, 
in fact, it agrees with the Poisson (small sky coverage) limit 
to order $\sigma^2$.  This agreement gives us the confidence to apply 
equation~(\ref{eqn:Pdelta}) to any relevant survey area.

\subsection{Projection along the line of sight}
\label{subsec:projection}
The SZ power spectrum is obtained by summing over all redshift bins. 
Since the contributions  from different redshift bins are correlated, 
the relation between the PDF of the total SZ power spectrum and the 
PDF of the SZ power spectrum from each redshift redshift bin is 
complicated.  However, for a sufficiently wide bin in $z$, 
e.g. $\Delta z=0.1$, the fluctuations $\delta_z$ of different 
redshift bins are approximately uncorrelated.   
In this case, the Fourier transform of the SZ power spectrum PDF is 
just the product of equation~(\ref{eqn:Pdelta}) for each redshift bin,  
\ba
\label{eqn:Fourier}
\tilde{P} = \prod_j \,\tilde{p}_j &=& \exp\left[\sum_j
  \left(D_jf_{\rm sky}+\frac{E_j^2\sigma_j^2f^2_{\rm sky}}{2}\right)\right]\\
&\equiv& \exp\Bigl[G(\tilde{f})\Bigr]   \ . \nonumber
\ea
Here, $j$ denotes the $j$-th redshift bin.

The Inverse Fourier transform of equation~(\ref{eqn:Fourier}) 
yields the PDF 
\be
 P(C_\ell) = \int \tilde{P}\exp(-{\rm i}fC_\ell)\,df/(2\pi).
\ee
Equation~(\ref{eqn:Fourier}) has all desired quantities: 
 (1) $\tilde{P}^{*}(f)=\tilde{P}(-f)$, so the PDF is real;
 (2) $\tilde{P}(f=0)=1$, so $\int P(C)dC=1$, meaning the PDF is correctly 
     normalized.  
It is straightforward to verify that 
$d\tilde{P}/df|_{f=0}={\rm i}\bar{C}$ and 
$d^2\tilde{P}/df^2=-(\bar{C}^2+\sigma_C^2)$ reduce to the 
expressions for $\bar{C}$ and $\sigma_C^2$ given in the previous 
section (equations~\ref{eqn:meanC} and \ref{eqn:variance} respectively). 
This shows explicitly that our expression for 
$P(C)$ has the correct first and second moments.

\subsection{2-point joint pdf of $C_\ell$ and $C_{\ell^{'}}$}
The joint pdf of $P_2(C_\ell,C_{\ell^{'}})$ can be derived 
similarly.  The 2D Fourier transform is 
\be
\label{eqn:2DFourier}
\tilde{P}_2(f,f^{'})\equiv \int\int
 P_2(C,C^{'})\,\exp\left({\rm i}fC+{\rm i}f^{'}C^{'}\right)\,df\,df^{'}\ .
\ee
For a single redshift, 
\ba
\tilde{P}_2 &=& 
  \int P_{\rm L}(\delta_1,\cdots,\delta_{\alpha})\,\prod_{\alpha}
                 \,d\delta_{\alpha}\\
   &&\ \times \quad\exp\left[\sum_{\alpha}
  N_{{\alpha},{\rm L}}\Bigl(e^{{\rm i}\tilde{f}S_{\alpha}+{\rm i}\tilde{f}^{'}S_{\alpha}^{'}}-1\Bigr)\right] \nonumber\ .
\ea
Again, assuming the linear bias model and Gaussian $p_j(\delta)$ yields 
\be
\label{eqn:Fourier2D}
\ln\tilde{P}_2 = 
  \sum_j f_{\rm sky}D_{2j} + \frac{f_{\rm sky}^2E_{2j}^2\,\sigma_j^2}{2}
  \equiv G_2(\tilde{f},\tilde{f}^{'}),
\ee
where 
\be
D_2 = f^{-1}_{\rm sky} \,
 \int d\bar{N}\left[e^{{\rm i}\tilde{f}S+{\rm i}\tilde{f}^{'}S^{'}}-1\right]\ ,
\ee
and 
\be
E_2 = f^{-1}_{\rm sky} \,
 \int d\bar{N}\,b\,\left[e^{{\rm i}\tilde{f}S+{\rm i}\tilde{f}^{'}S^{'}}-1\right]\ .
\ee 
It is straightforward to verify that this expression is 
normalized to unity, and that the expressions for
 $\bar{C},\bar{C}^{'},\sigma_C,\sigma_{C^{'}}$ and 
${\rm Cov}_{\ell\ell^{'}}$ which it implies are all in agreement 
with our previous expressions (e.g. equation~\ref{eqn:covariance}).

If $G_2(\tilde{f},\tilde{f}^{'})$ can be written as the sum of two 
terms, one a function of $\tilde{f}$ and the other of $\tilde{f}^{'}$, 
then this would imply that $C$ and $C'$ are independent.  The 
expression above shows that, in general, this condition is not 
satisfied. For example, there are cross terms in both $D$ and $E^2$ 
which imply correlations between $C$ and $C^{'}$.  
In the limit that $\ell$ and $\ell^{'}$ are sufficiently different 
that $S\gg S^{'}$ (or $S\ll S^{'}$), then $D_2$ roughly meets this 
condition.  Since the correlation induced by the $E^2_2$ term is
weak, the resulting $P_2(C,C^{'})\simeq P(C)P(C^{'})$.

\subsection{$n$-point joint pdf of $C_\ell$ and $C_{\ell^{'}}$}
The analysis above for the Fourier Transform of the 2-point pdf 
is easily generalized to that for the $n$-point pdf: 
\ba
 \label{eqn:Fourier_n_defination}
  \ln\tilde{P}_n(f_1,\cdots,f_n)&\equiv &
   \ln\int\int P_n(C_1,\cdots,C_n) \nonumber\\
 &&\quad\times\quad 
  {\rm e}^{{\rm i}f_1C_1+\cdots+{\rm i}f_nC_n}\,df_1\cdots df_n
   \nonumber \\
 &=&\sum_j f_{\rm sky}D_{nj} + \frac{f^2_{\rm sky}E_{nj}^2\,\sigma_j^2}{2},
 \label{eqn:Fourier_n}
\ea
where
\be
D_{n} = f_{\rm sky}^{-1}\,\int d\bar{N}\,
\left[{\rm e}^{{\rm i}\tilde{f}_1S_1+\cdots+{\rm i}\tilde{f}_nS_n}-1\right]\ ,
\ee
and 
\be
E_{n}=f_{\rm sky}^{-1}
 \int d\bar{N}\,b\,
 \left[{\rm e}^{{\rm i}\tilde{f}_1S_1+\cdots+{\rm i}\tilde{f}_nS_n}-1\right]\ .
\ee
Equation~(\ref{eqn:Fourier_n}) for the $n$-point PDF and its special 
cases, equations~(\ref{eqn:Fourier}) and~(\ref{eqn:Fourier2D}) for 
the 1- and 2-point PDFs, are the key results of this paper.


\bfi{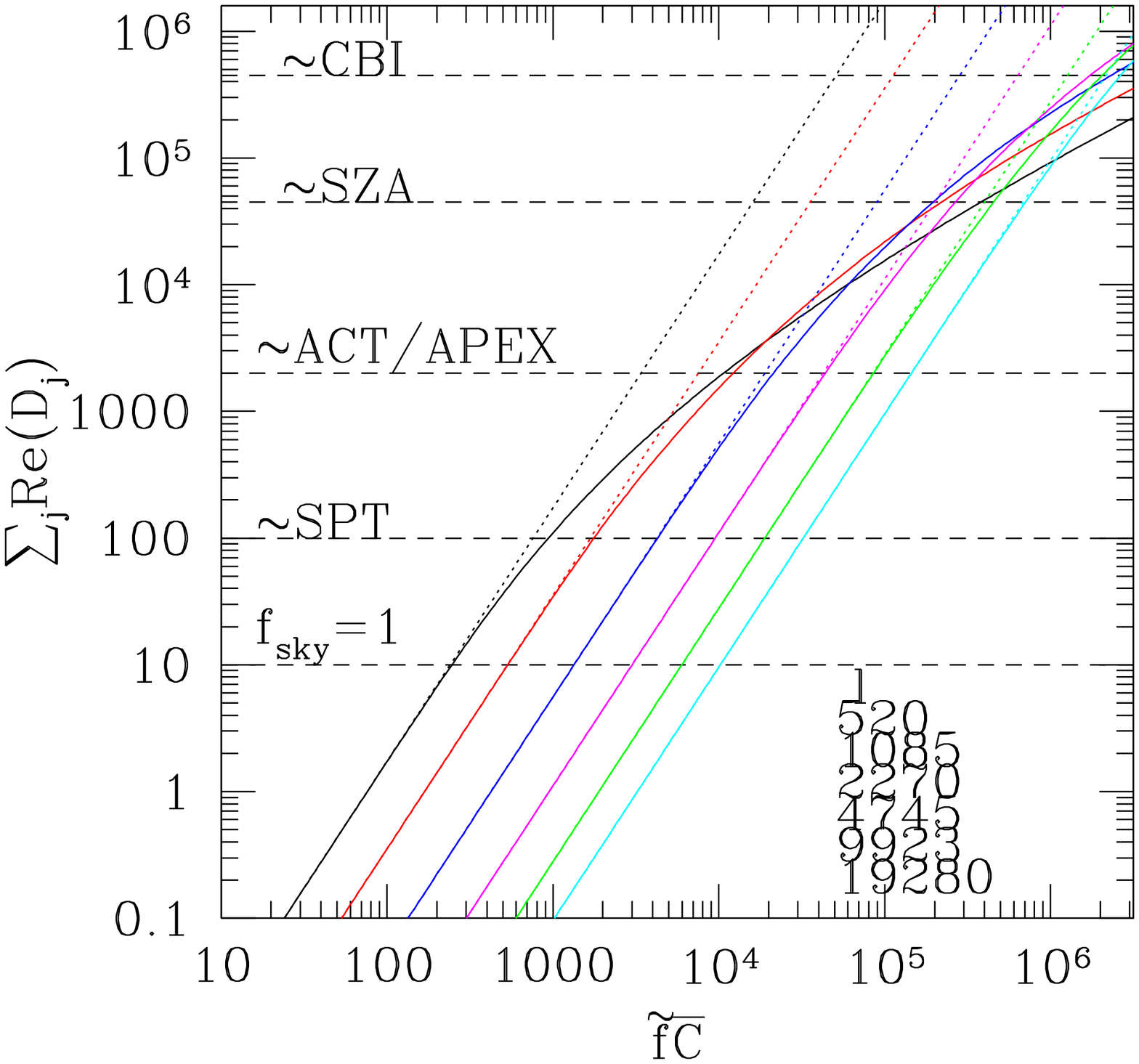}
\caption{The requirement for having a Gaussian PDF. 
  Solid lines show ${\rm Re}(\sum_j D_j)$ for several multipoles. 
  Straight dotted lines are $\propto f^2$. 
  The deviation of the dotted lines from the corresponding $f$ 
  modes implies that the pdf is non-Gaussian.  
  The $\tilde{f}$ modes below the straight dashed lines are the 
  relevant modes for corresponding sky coverage (SPT: $\sim 4000$
  deg$^2$; ACT/APEX: $\sim 200$ deg$^2$; SZA: $\sim 10$ deg$^2$ and
  CBI: $\sim 1$ deg$^2$). 
  \label{fig:check}}
\efi


\subsection{The large sky coverage limit}
To better understand the physical meaning of the terms in
equations~(\ref{eqn:Fourier}), (\ref{eqn:Fourier2D}) and 
(\ref{eqn:Fourier_n}), it is useful to study the limit of large 
sky coverage.

Consider the one point pdf $P(C)$ in this limit. 
Effectively, only those modes with ${\rm Re}[G(\tilde{f})]\ga -10$ 
contribute to $P$.  Over this range of $\tilde{f}$, 
 ${\rm Re}[G]$ decreases with $\tilde{f}$. 
Thus, those modes with ${\rm Re}[G]\ga -10$ have 
 $\tilde{f}\la \tilde{f}_c$, where ${\rm Re}[G(\tilde{f}_c)]=-10$. 
Since $|{\rm Re}(G)|$ increases with $f_{\rm sky}$, 
$\tilde{f}_c$ decreases with $f_{\rm sky}$. 
Thus, for sufficiently large $f_{\rm sky}$, $\tilde{f}_c$ and so 
all relevant $\tilde{f}$ are small, and 
$\exp({\rm i}\tilde{f}C)-1\simeq {\rm i}\tilde{f}C - (\tilde{f}C)^2/2$. 
In this case, 
\ba
\label{eqn:approximations}
 {\rm Im}(D_j)&\simeq& \tilde{f}\bar{C}_j,\qquad 
 {\rm Re}(D_j)\simeq-\frac{\tilde{f}^2}{2}\frac{\bar{C}_j^2}{\bar{N}_j},
 \nonumber\\ 
 {\rm Im}(E_j)&\simeq& \tilde{f} \bar{b}_j\bar{C}_j,\qquad
 {\rm Re}(E_j)\simeq 0, \quad {\rm and} \\\nonumber
 G(\tilde{f})&\simeq&  {\rm i}f\bar{C}-\frac{f^2}{2}
  \sum_j \bar{C}_j^2\left[\frac{1}{N_{{\rm P}j}f_{\rm sky}}
                            +\bar{b}_j^2\sigma_j^2\right]\nonumber \\
             &=& {\rm i}f\bar{C}-\frac{f^2\sigma_C^2}{2}\ .
\label{eqn:G}
\ea
Here, 
\ba
 \bar{C}_{j} &\equiv& f^{-1}_{\rm sky}\int S(\lambda)d\bar{N}_j,\ \ 
 N_{{\rm P}j} \equiv  f^{-1}_{\rm sky}
        {\left[\int S(\lambda)\,d\bar{N}_j\right]^2\over 
               \int S(\lambda)^2\,d\bar{N}_j} \ \ {\rm and}\nonumber\\
 \bar{b}_{j} &\equiv& {\int S(\lambda)b(\lambda)\,d\bar{N}_j\over 
                          \int S(\lambda)\,d\bar{N}_j}
\ea
are all defined in the $j$-th redshift bin. 
$\bar{C}$ is defined by equation~(\ref{eqn:meanC}) and 
$\bar{C}=\sum \bar{C}_j$.
None of these quantities depends on $f_{\rm sky}$.  
Notice that the coefficient of $f^2/2$ in equation~(\ref{eqn:G}) 
is just the variance of $C$ (equation~\ref{eqn:variance}).  
Thus, in the large sky limit, there are sufficiently many clusters 
that the Poisson distribution becomes Gaussian, in accordance with 
the central limit theorem.  Thus $P$, the Fourier transform of 
$\exp[G]$,  will also be Gaussian.

When the above conditions are satisfied, the Fourier transforms of 
equations~(\ref{eqn:Fourier}) and~(\ref{eqn:G}) yield 
\be
\label{eqn:Gaussian}
 P(C) = \frac{1}{\sqrt{2\pi}\sigma_C}
        \exp\left(-\frac{(C-\bar{C})^2}{2\sigma_C^2}\right),
\ee
where $\bar{C}$ and $\sigma_C$ are given by 
equations~(\ref{eqn:meanC}) and~(\ref{eqn:variance}), respectively. 
This shows that, in the limit of large sky coverage, $P(C)$ becomes
Gaussian.

However, the conditions ({\ref{eqn:approximations}) are not always 
satisfied.  The condition 
${\rm Re}(D_j)\simeq (f/f_{\rm sky})^2\bar{C}_j^2/\bar{N}_j$ 
is especially hard to meet.  Since the term ${\rm Re}(\sum_j D_j)$ 
dominates that involving $\sum_j E^2_j$ (consistent with 
Fig. \ref{fig:variance}), requiring that
\be
 \sum_j {\rm Re}(D_j)\propto f^2 \quad {\rm when}\quad
 f_{\rm sky}\sum_j {\rm Re}(D_j)\geq -10 ,
 \label{eqn:gausscond}
\ee
is a suitable condition for Gaussianity.  
We call this the {\it Gaussian condition}. 
Fig. \ref{fig:check} shows that, for an all sky survey, all 
$\ell > 500$ satisfy the Gaussian condition given above. 
For SPT,  the Gaussian condition is satisfied at 
$\ell\ga 10^3$. However, for small sky coverage survey such as SZA, 
the Gaussian condition is only satisfied at $\ell\ga 10^4$.

In the limit of large sky coverage, one can Taylor expand $G_2$ 
around $\tilde{f}, \tilde{f}^{'}=0$, keeping terms up to $\tilde{f}^2,
\tilde{f}^{'2}$. The resulting 2-point joint PDF takes the Gaussian 
form
\be
 P_2\propto \exp\left[-\frac{X^2+X^{'2}-2XX^{'}r}{2(1-r^2)}\right],\ {\rm
   where}\ X=\frac{C-\bar{C}}{\sigma_C} 
\ee
and $r$ is given by equation~(\ref{eqn:r}). 
If $\ell^{'}\sim \ell$, then $C_\ell$ and $C_{\ell^{'}}$ are 
dominated by contributions from halos with similar $M$, $z$ and 
$\lambda_\ell$. So we expect $C_\ell$ and $C_{\ell^{'}}$ to be 
highly correlated. Their correlation is quantified by $r$ and as 
expected, when $\ell^{'}\to \ell$, 
$r\rightarrow 1$. When $\ell^{'}$ differs significantly from 
$\ell$, $C_\ell$ and $C_{\ell^{'}}$ are dominated by different 
clusters, so $r\to 0$, meaning $C_l$ and $C_{l^{'}}$ are nearly 
independent.


\bfi{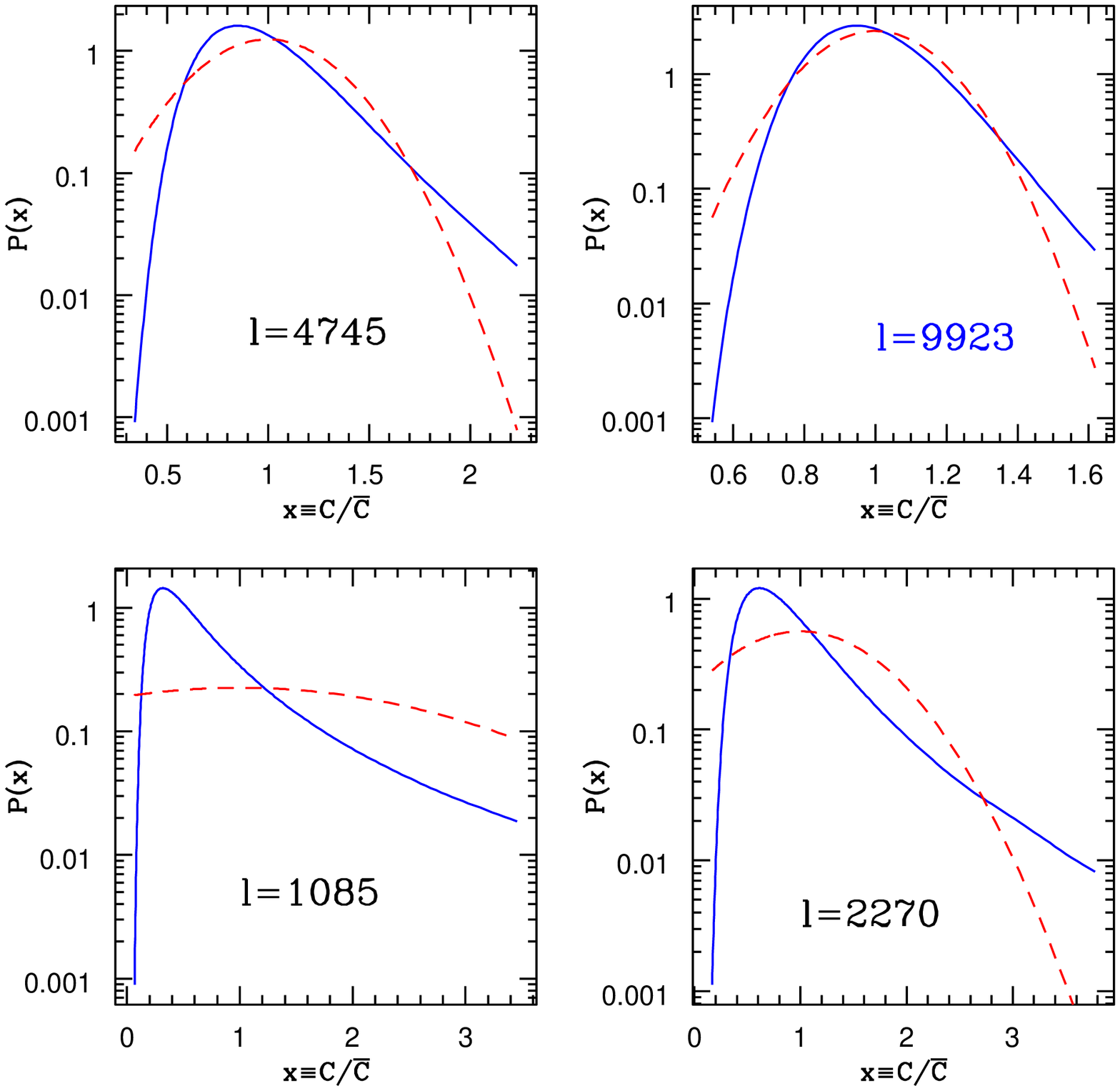}
\caption{The PDF of $C\equiv C_\ell$ for $\ell=1085,2270,4745,9923$,
  respectively in a $1$ deg$^2$ survey. Solid lines show the pdf 
  given by equation~(\ref{eqn:Fourier}), and dashed lines show a 
  Gaussian form with the same mean $\bar{C}$ and $\sigma_C$ for 
  comparison. 
  $P(C)$ is highly non-Gaussian for $\ell\la 4000$, mainly because 
  the small survey volume means that discreteness fluctuations 
  in the cluster counts are large (c.f. Fig. \ref{fig:check}). 
  \label{fig:1}}  
\efi

\subsection{Numerical evaluation of the PDF}
\label{sec:numerical}
The integral form of the $n$-point SZ power spectrum PDF
(equation~\ref{eqn:Fourier_n}) must be computed numerically.  
This is straightforward for the 1- and 2-point pdfs, but more 
sophisticated Monte-Carlo methods must be used for efficient 
evaluation when $n$ is large.  In what follows, we illustrate 
our results using the 1-point pdf (equation~\ref{eqn:Fourier}).

The solid curves in Figs. \ref{fig:1}, \ref{fig:10} and \ref{fig:200} 
show equation~(\ref{eqn:Fourier}) for surveys covering $1$, $10$, 
$200$ and $4000$ deg$^2$.  For comparison, we also show the 
corresponding Gaussian form (equation~\ref{eqn:Gaussian}).  
For $1$ deg$^2$ sky coverage, representative of  CBI, $P(C)$ is highly
non-Gaussian at $\ell\la 4000$. This is consistent with what would be
expected from Fig. \ref{fig:check}, which shows that a significant 
fraction of the relevant modes deviate from the 
${\rm Re}\sum D_j\propto f^2$ scaling, so the Gaussian condition 
(equation~\ref{eqn:gausscond}) is not satisfied.  
Further understanding of this non-Gaussian behavior comes from 
Fig. \ref{fig:variance}:  the effective number of clusters contributing 
to the SZ effect is $N_{\rm P}\,f_{\rm sky} \la 10$ for $\ell<4000$. 
So we expect large Poisson fluctuations in the SZ power spectrum 
for $\ell < 4000$.  Although $N_p$ increases with $\ell$, making the 
non-Gaussianity weaker at higher $\ell$, it is still only $\sim 30$ 
at $\ell=10^4$, so we expect $P(C)$ to be mildly non-Gaussian even 
at $l=10^4$. 
For $10$ deg$^2$ sky coverage, roughly the area 
SZA\footnote{SZA, http://astro.uchicago.edu/sza/} plans to cover, 
$P(C)$ is strongly skewed at $l\la 2000$, with a non-negligible tail 
at high $C$. Only for multipoles $\ga 5000$, where 
$N_pf_{\rm sky}\ga 100$ (Fig. \ref{fig:variance}), does $P(C)$ 
approach the Gaussian form (Fig. \ref{fig:10}). 
For larger sky coverage, $P(C)$ becomes more Gaussian. 
For a $200$ deg$^2$ survey, roughly 
ACT\footnote{ACT,http://www.physics.princeton.edu/act/} and 
APEX\footnote{APEX,http://bolo.berkeley.edu/apexsz/index.html}
 plans to cover, $P(C)$ approaches Gaussian at $l\ga 2000$.  
Note, however, analysis of the SZ power spectrum at $\ell\la 2000$ must 
account for the fact that the pdf is non-Gaussian.  
Again, these results are consistent with expectations based on 
Figs. \ref{fig:variance} and \ref{fig:check}.


\bfi{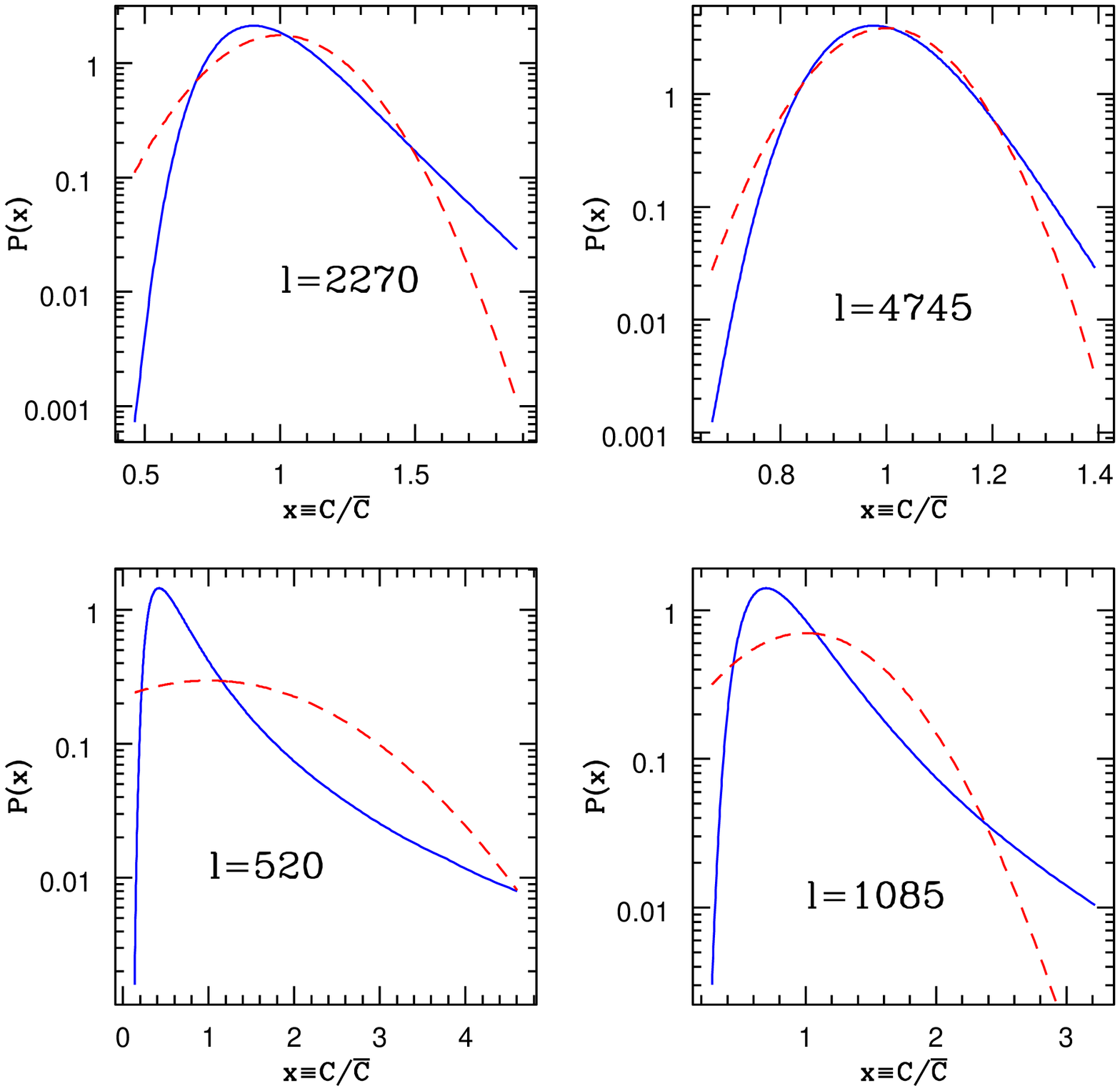}
\caption{Similar to Fig. \ref{fig:1}, but for $10$ deg$^2$ sky
  coverage and $\ell=520,1085,2270,4745$, respectively.  The 
  non-Gaussianity is weaker compared to that in Fig. \ref{fig:1}, 
  because the larger number of clusters in the survey volume brings
  the Poisson fluctuations closer to the Gaussian limit. \label{fig:10}} 
\efi

For $4000$ deg$^2$ sky coverage, roughly the size of 
SPT\footnote{SPT, http://spt.uchicago.edu/}, $P(C)$ is Gaussian at 
$\ell\ga 1000$.  For Planck, which will cover the full sky, $P(C)$ 
can be assumed to be Gaussian distributed at all relevant scales. 
While this will significantly simplify the analysis of the SPT and 
Planck SZ power spectra, we caution that the variance of this 
Gaussian differs significantly from that expected in a truly 
Gaussian random field (fig. \ref{fig:variance}).  The 
strong correlation between different $\ell$ modes makes the total 
S/N of the SZ power spectrum significantly smaller than that of a 
random Gaussian field.

In our model,  the SZ power spectrum is dominanted by contributions
from single halos (the one-halo term), and we have attributed the
non-Gaussianity in the SZ $p(C_\ell)$ we find to Poisson fluctuations
in the halo  number distribution.
In this model, different $l$ modes of the  SZ effect can be highly
correlated, with correlation length $\Delta\ell_c$ comparable to $\ell$
(Fig. \ref{fig:r}).
Hence, the PDF of the band power with width $\Delta\ell$ is roughly
the same as the PDF of a single mode, as long as
$\Delta\ell\leq \Delta\ell_c$,
The strength of  the non-Gaussianity in this band power PDF is then
also determined by the effective number of halos in the survey volume,
which is $N_pf_{\rm sky}$.

However, there is another possible origin of the non-Gaussian PDF.
Suppose that the distribution of SZ temperature fluctuations were
exactly Gaussian.  Then $p(C_\ell)$ would follow a
$\chi^2$-distribution, so one might wonder if this is the origin of
the non-Gaussianity seen in Figs.~\ref{fig:1}, \ref{fig:10} and
\ref{fig:200}?  The argument for band-powers, rather than single
modes, is similar:  If $Y_i$ is the $i$-th Fourier mode, and we
define the normalized band power $x = \sum_i^n |Y_i|^2/(n\bar{C})$,
where $\bar{C}$ is the mean variance of the $Y_i$, then $x$ is the
average of $n$ independent Fourier modes.  If we approximate the
variance in each mode as being the same as $\bar C$, then
\be
 P_n(x) = \frac{2^{-n}}{\Gamma(n)}\, x^{n-1}\,\exp(-x/2)\,2n\ .
\ee
is a $\chi^2$-distribution with $2n$ degrees of freedom
(recall each $Y_i$ has independent real {\em and} imaginary parts).
Note that $P_n(x)$ becomes increasingly Gaussian in shape as $n\gg 1$,
but it is very non-Gaussian for small $n$.

For a survey with fractional sky coverage $f_{\rm sky}$,
$n = \ell\Delta\ell\, f_{\rm sky}$, where setting the bin size
$\Delta\ell$ equal to the correlation scale $\Delta\ell_c$ is a
natural choice.
Hence, if $n = \ell\Delta\ell_c\, f_{\rm sky}\gg N_p\, f_{\rm sky}$,
then the non-Gaussianity in the band power PDF will be mainly
contributed by Poisson fluctuations in halo number.
Now, Fig. \ref{fig:r} shows that $\Delta\ell_c\sim \ell$,
and Fig. \ref{fig:variance} shows that, at $\ell\sim 10^3$,
$N_{\rm P}\approx (\ell/10)^2$, so
$n\approx\ell^2$ is indeed larger than $N_p\approx (\ell/10)^2$.
This shows that the non-Gaussian shape of $p(C_\ell)$ in our model
is dominated by the Poisson fluctuations in halo (cluster) counts;
Fig.~\ref{fig:origin} shows this explicitly.


\bfi{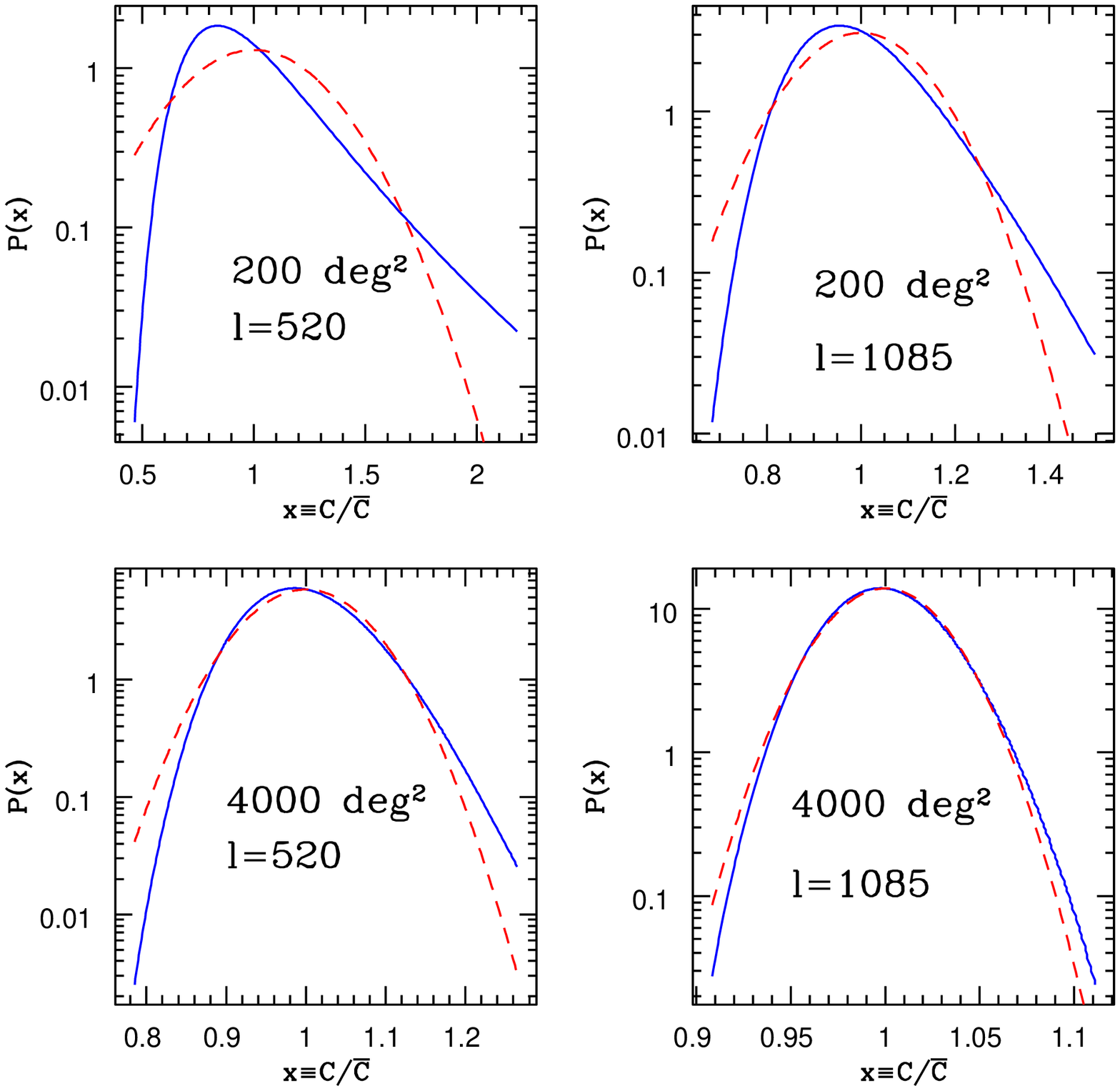}
\caption{Similar to Fig. \ref{fig:10}, but for $200$ and 
  $4000$ deg$^2$ sky coverage and $\ell=520, 1085$, respectively. 
  Comparison with the smaller surveys shown previously ($1$ and 
  $10$ deg$^2$) shows that the pdf becomes increasing Gaussian 
  as the sky coverage increases. \label{fig:200}}   
\efi

\bfi{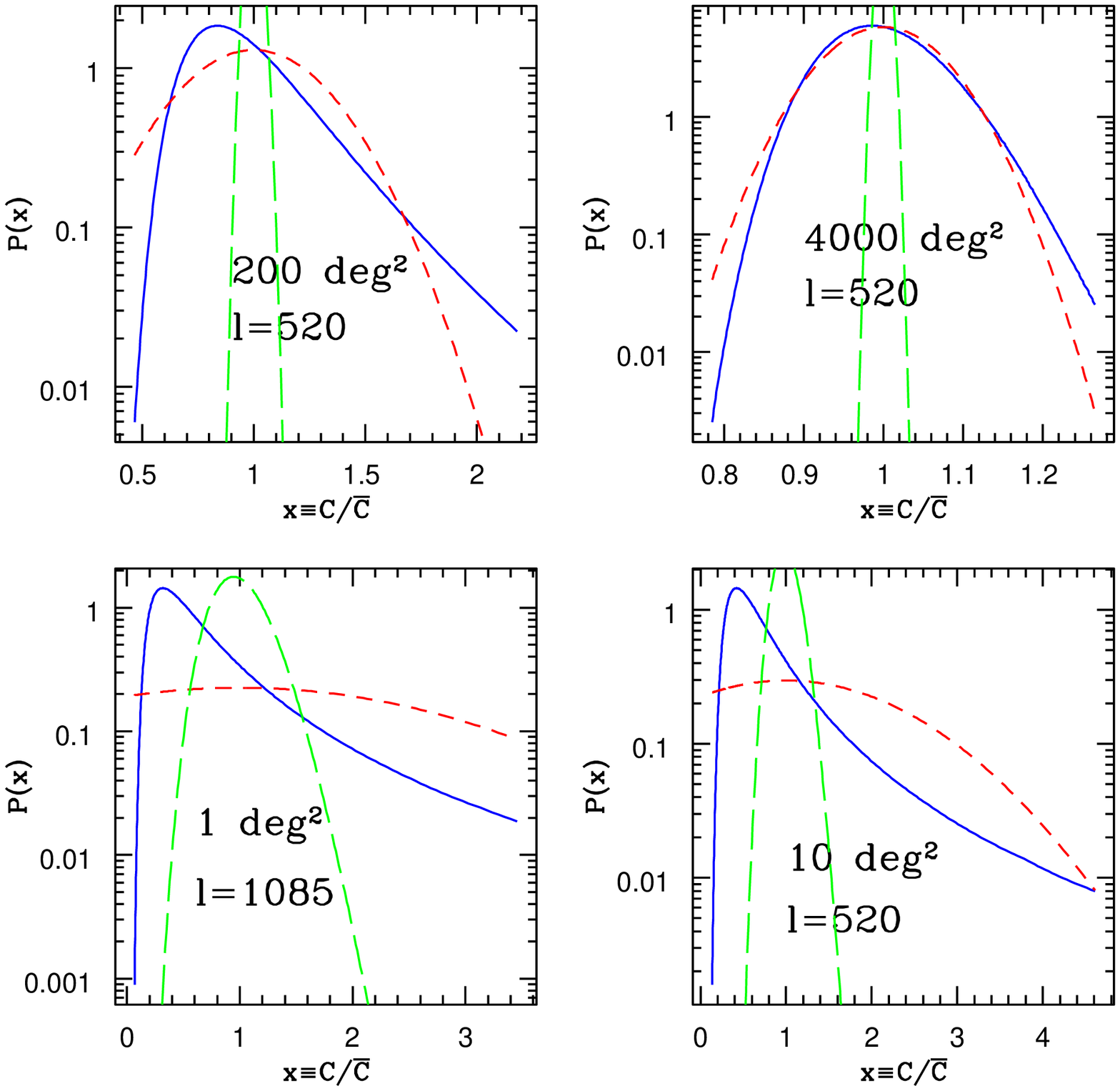}
\caption{The origin of the  non-Guassianity in the SZ band power
  spectrum PDF.
  If the SZ signal were Gaussian distributed, then the PDF of the
  SZ band-power would follow a $\chi^2$-distribution with
  $2\ell \Delta \ell  f_{\rm sky}$ degrees of freedom, and one might
  wonder if this was the origin of the non-Gaussian shape of the
  SZ PDF.  To illustrate that this is not the origin of the
  non-Gaussianity in our model, the long-dashed lines show
  Chi-square distributions when $(\ell,\Delta\ell)=(520,400)$
  and $(1085,800)$, respectively.
  They are clearly different from the solid curves which show the
  PDF predicted by our model; short dash lines show a Gaussian form
  with the same rms.  (In fact, because close $\ell$ modes are
  strongly correlated, the true PDF of the SZ band power with bin
  size $\Delta \ell \leq \Delta \ell_c$ is roughly the same as the PDF of
  a single Fourier mode (Fig. \ref{fig:1}, \ref{fig:10} and
  \ref{fig:200}), so we have simply copied the corresponding curves
  here.)
  \label{fig:origin}}
\efi

The PDF of the SZ power spectrum  is similar to that expected 
from gravitational lensing in several respects:
 (1) $P(C)$ is skewed so that it peaks at $C<\bar{C}$;
 (2) It has a long tail extending to high $C$.
For BIMA/CBI, this increases the probability of having $C\geq 2\bar{C}$ 
at the relevant $\ell$ range.  When combined with the large sample 
variance caused by strong correlations between different $\ell$ 
modes, this may help reduce the tension between the CBI/BIMA power 
excess relative to the naive expectations based on primary CMB and 
the Gaussian model of $P(C)$.  
We postpone a quantitative discussion of this issue to elsewhere.

However, the angular scale dependence of the SZ power spectrum PDF 
is distinctively different from that due to gravitational lensing. 
The lensing $C_\ell$ is dominated by the two-halo term at large 
scales, whereas the SZ power spectrum is always dominated by the 
one-halo term \citep{Komatsu99}.  Hence the pdf induced by lensing 
is expected to become Gaussian at large angular scales (small $\ell$), 
whereas the SZ PDF becomes Gaussian at small angular scales 
(large $\ell$) (\citet{Zhang01} and see Figs. \ref{fig:1}, 
\ref{fig:10}, and \ref{fig:200}).  
Furthermore, the large scale SZ power is mainly contributed by massive 
clusters, whose number density is low (e.g. \citet{Komatsu02} and 
Fig. \ref{fig:variance}) . This causes large Poisson fluctuations and 
strong deviation from Gaussianity at large scales. On smaller scales, 
where more and more small clusters contribute, the Poisson 
fluctuations tend to the Gaussian limit.

A remaining question is the dependence of $p(C_\ell)$ on 
cosmological parameters, especially $\sigma_8$.  
A convenient measure of $p(C_\ell)$ is the effective number of 
clusters $N_{\rm P}$ (equation~\ref{defNp}).  
There are essentially two effects:  
(1) A change in $\sigma_8$ changes the concentrations of halos, 
    and hence relative weights of clusters. Since $N_{\rm P}$ is the number of
    clusters weighted by their SZ contributions, change in $c$  affects
    $N_{\rm P}$.   
   (i) Roughly, a cluster contributes to the SZ power spectrum 
     only at scales larger than $r_s=r_{\rm vir}/c\propto M^{1/3}/c$; 
     the SZ signal from a cluster is smooth on scales smaller than $r_s$.
     Here, $c$ is the concentration parameter and $r_{\rm vir}$ is 
     the virial radius of a cluster of mass $M$. 
     Since $c$ decreases with increasing mass, massive clusters 
     only contribute to the SZ signal on relatively large scales; 
     whereas lower mass clusters contribute down to smaller scales.    
     Now, increasing $\sigma_8$ makes clusters of a fixed mass more 
     concentrated---at fixed mass $r_s$ decreases. 
     Thus, a massive cluster which does not contribute power to a 
     given scale if $\sigma_8$ is low may be able to contribute at 
     higher $\sigma_8$.  Thus, increasing $\sigma_8$ increases the 
     contribution from rarer more massive clusters.  
  (ii) Increasing $\sigma_8$ is expected to increase $c$ by the same 
     fraction for all masses \citep{Bullock01,Wechsler02,Zhao03}. 
     However, a constant fractional increase in $c$ results in 
     larger fractional increase in the SZ flux from a more massive 
     cluster. Thus the weighting of more massive clusters in the SZ
     power spectrum is increased when increasing $\sigma_8$.  
(2) Increasing $\sigma_8$ increases the number density of massive 
    clusters, making $N_{\rm P}$ larger and so Poisson fluctuations 
    in cluster abundances smaller.

Effect (1) means that increasing $\sigma_8$ increases the relative 
contribution of the rarer more massive clusters (by increasing $c$); 
if cluster abundances were not also altered, this would decrease 
$N_{\rm P}$ (since massive clusters are rarer).  However, increasing 
$\sigma_8$ also increases the number of more massive clusters (effect 2), 
so the two effects act in approximately opposite directions.  
Figure~\ref{fig:NpSigma8} shows the result of an explicit calculation 
of how $N_{\rm P}$ depends on $\sigma_8$:  the two effects do indeed 
approximately cancel.


\bfi{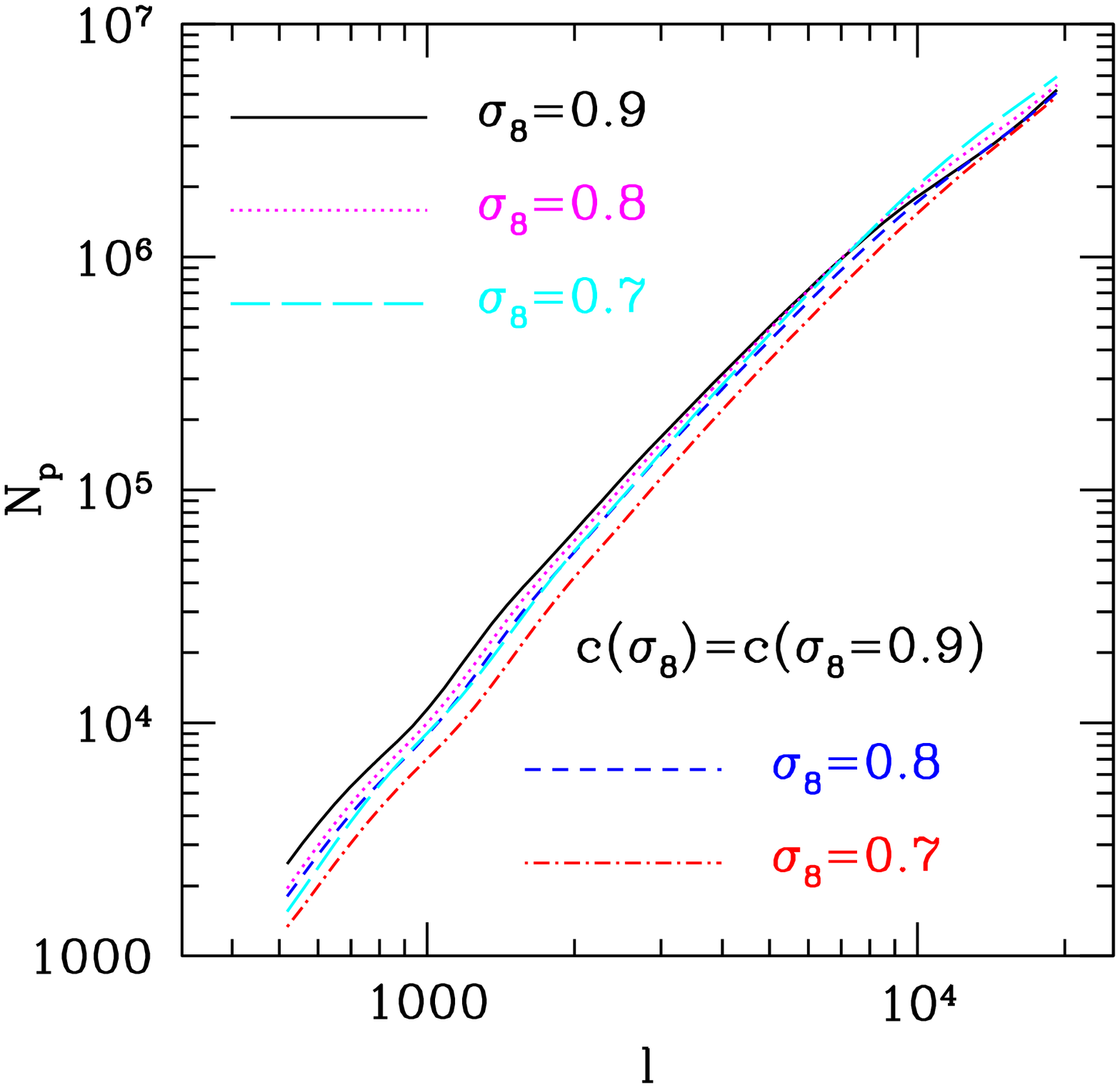}
\caption{Dependence of the effective number of clusters $N_{\rm P}$ 
         (equation~\ref{defNp}) on $\sigma_8$.  
         The short-dashed ($\sigma_8=0.8$) and dot-dashed lines 
         ($\sigma_8=0.7$) have the same relation between concentration 
         $c$ and halo mass $m$ as when $\sigma_8=0.9$.  
         Hence, the difference in $N_{\rm P}$ is entirely due to the 
         change in the mass function $dn/dm$.  
         At large $\ell$, where less massive clusters dominate, the 
         dependence on $\sigma_8$ is negligible.  This is because 
         the abundance of the most massive clusters is exponentially 
         sensitive to $\sigma_8$, whereas less massive clusters are 
         less strongly dependent on $\sigma_8$.  
         The dotted ($\sigma_8=0.8$) and long-dashed ($\sigma_8=0.7$) lines account for the additional 
         fact that halo concentrations also dependend on $\sigma_8$.  
         This dependence partly cancels that from $dn/dm$. 
         At small $\ell$, the effect on $dn/dm$ dominates; 
         at large $\ell$, the effect on $c$ dominates. 
         Hence, at $\ell\sim 10^4$, $N_{\rm P}$ when $\sigma_8=0.7$ 
         is larger than when $\sigma_8=0.9$.  }
\label{fig:NpSigma8}   
\efi

When expressed as a function of $x\equiv C_{\ell}/\bar{C}_\ell$, 
the PDF $p(C_\ell)$ is approximately invariant to changes in $\sigma_8$. 
This provides some analytic support for the procedure of 
\citet{Dawson06} which we discussed in the Introduction. 
However, our analysis indicates that this procedure is only  
approximately correct.  In reality, effects (1) and (2) do not 
cancel exactly; we do see differences in $N_p$ and $p(C_\ell)$ as 
$\sigma_8$ is varied. 
Robust evaluation of the net effect requires improvements to the 
models we use for cluster abundance, the $c-M$ relation and the gas 
model, as well as the incorporation of several complexities which 
we discuss in the next section.  These improvements will be addressed 
elsewhere.

\section{Summary, discussion of limitations, and extensions}
\label{sec:summary}\label{sec:discussion}
We have derived an analytic estimate of the $n$-point PDF of the 
SZ power spectrum (equation~\ref{eqn:Fourier_n}). 
Our derivation is based on the halo-model.  The analytical integral 
form of the PDF allows fast calculation of 
the one- and two-point pdfs, $P(C_\ell)$ and $P(C_\ell,C_{\ell'})$ 
(equations~\ref{eqn:Fourier} and~\ref{eqn:Fourier2D}). 
More advanced integration routines are required to compute 
higher-order PDFs in reasonable timeframes.

We find that the non-Gaussianity of the SZ power spectrum is a 
function of scale $\ell$, and, at fixed  $\ell$, is a strong function 
of the survey area (Figs. \ref{fig:1}, \ref{fig:10} and \ref{fig:200}).   
This strong non-Gaussianity may resolve the discrepancy between 
CBI/BIMA SZ measurements and the primary CMB measurements.
For survey areas $\sim 10$ deg$^2$, which will be achieved by 
ongoing surveys such as SZA, the non-Gaussianity is 
significant at $\ell\la 4000$, but becomes negligible at $\ell\ga 5000$. 
For future SZ surveys such as SPT, which will cover $10\%$ of the sky, 
the PDF of the SZ power spectrum can be approximated  as Gaussian at 
$\ell\ga 1000$. For Planck, the PDF of the SZ power spectrum can be 
approximated as Gaussian at all relevant scales.

While useful, our results should be treated with caution for a 
number of reasons.  
(1) We have assumed cluster dark matter density profiles 
    are universal, and can be described by the  NFW profile.  
    N-body simulations show that a non-negligible fraction of 
    clusters are not well fit by this form \citep{Jing00}. 
(2) At fixed mass, halos which are well fit by the NFW form have 
    a range of concetrations \citep{Bullock01,Jing02}.  
    We have not accounted for this scatter.
(3) We have assumed that clusters are spherical. 
    In reality, clusters are better described as tri-axial 
    spheroids \citep{Jing02}. This will cause the SZ signal of a 
    cluster to be non-spherical \citep{Wang04}.  
These complexities will increase the variance of the distribution 
of $C_\ell$.

Of perhaps more concern is that fact that our assumption of a 
spherically symmetric profile leads to the unrealistic prediction 
that different ${\bf \ell}$ with the same amplitude $|{\bf \ell}|$ 
are completely correlated.  On the other hand, the measured 
quantity is almost always a band-power averaged SZ power spectrum, 
where the average is over different ${\bf \ell}$, so the assumption 
of spherical profiles should be accurate to lowest order.   
Nevertheless, it might be interesting to account for halo triaxiality.

Incorporating (1), (2) and (3) into our model 
(equations~\ref{eqn:Fourier} and~\ref{eqn:Fourier_n}) is relatively 
straightforward.  \citet{Cooray02} shows how the scatter in halo 
concentrations can be incorporated into the halo model, and 
\citet{Smith06} show how to incorporate the effects of halo 
triaxiality into models of the dark matter power spectrum.  
Since this is particularly straightforward for the one-halo term, 
it might be interesting to combine their analyses with ours.  
But doing so is beyond the scope of our present work.

While $P(C)$ is important for assigning realistic errors to 
measurements of the SZ power spectrum, thus permitting accurate 
constraints on cosmological paramteres, it also contains important 
information about cosmological and gas physics. 
For example, the shape of the tail at $C>\bar{C}$ is sensitive 
to the number of clusters in the survey and to the gas fraction 
in clusters.  With sufficiently large sky coverage, one can 
divide the survey area into many $1$ deg$^2$ patches, thus providing 
a direct measurement of the SZ PDF on $1$ deg$^2$ scales.
Comparison with Fig. \ref{fig:1} then constrains cluster abundances
and gas physics.   A detailed analysis of how much additional 
information (i.e. more than is provided by the power spectrum 
itself) is contained in the full PDF is beyond the scope
of this paper, although it certainly deserves further investigation.

It would have been nice to test our predictions with measurements 
from numerical simulations. However, current SZ simulations lack 
sufficiently large independent simulation volumes to measure the 
PDF reliably beyond the lowest moments. 
That said, we stress that the variance of the SZ power spectrum 
which our approach predicts is in excellent agreement with 
simulations (c.f. \S \ref{sec:variance}). 
This gives us confidence that our approach should at least provide 
a useful template for more accurate model of the SZ PDF.

The  analytical approach described in this paper can be applied to
several other cases. For example, it can be applied to calculate the
PDF of the lensing power spectrum at high $\ell$ straightforwardly, 
where the one halo term dominates. At low $\ell$ where two halo term 
becomes important or even dominant in the lensing power spectrum, 
this is less straightforward. Extending this approach to include 
the two halo term and so make predictions for the PDF of the weak 
lensing power spectrum is under investigation.

\section*{Acknowledgments}
We  thank Houjun Mo for  helpful discussions on the
application of the PDF to constrain cosmology and gas physics and
Oliver Zahn for useful comments. We thank the
anonymous referee for useful suggestions. PJZ thanks
the University of Pennsylvania  astrophysics group for the hospitality where
this work was initiated and  finalized.  
PJZ is supported  by the One-Hundred-Talent Program of the 
Chinese Academy of Science and the National Science Foundation of China grants
(No. 10533030,  10543004,10673022). RKS was supported by NASA-ADP grant
NNG05GK81G.

\appendix


\section{Small sky coverage limit}
\label{sec:appendix}
For surveys with very small sky coverage,  $|\exp[\int (e^{ifC}-1)dN]|\ll 1$. 
One can then Taylor expand 
\ba
 \exp\int (e^{ifC}-1)dN\simeq 1 + \int (e^{{\rm i}\tilde{f}S}-1)dN 
   + {[\int (e^{{\rm i}\tilde{f}S}-1)dN]^2\over 2}. \nonumber
\ea
Inserting into equation~(\ref{eqn:biasP}) and Fourier transforming 
to real space gives $P(C)$ in the Poisson limit:  
\ba
\label{eqn:small_onebin}
P(C)&=&(1-\bar{N})\delta_D(C)+\int \delta_D(Cf_{\rm sky}-S)d\bar{N} \nonumber\\
    &+ &\frac{1}{2}\left(\int \delta_D(S_1+S_2-Cf_{\rm
    sky})d\bar{N}_1d\bar{N}_2 \right. \nonumber \\
&&\left.\ \  -2[\int
   \delta_D(Cf_{\rm sky}-S)d\bar{N}] \bar{N}+\bar{N}^2\delta_D(C)\right)
\nonumber \\
&+& \frac{\sigma^2}{2}\left(\int
  \delta_D(S_1+S_2-Cf_{\rm sky})b_1b_2d\bar{N}_1d\bar{N}_2 \right. \nonumber
    \\
&&\left. - 2[\int 
   \delta_D(Cf_{\rm sky}-S)bd\bar{N}]\int bd\bar{N}+[\int
 bd\bar{N}]^2\delta_D(C)\right) \nonumber
\ea
Here, $\bar{N}=\int d\bar{N}$.  
Note that this expression does not rely on the assumption that 
$P_\delta$ is Gaussian. One can Taylor expand equation~(\ref{eqn:Pdelta})
in the limit $f_{\rm sky}\rightarrow 0$, perform the Fourier Transform, 
and verify that equation~(\ref{eqn:Pdelta}) does indeed agree with 
the above result to order $O(\sigma^2)$.  This suggests that
equation~(\ref{eqn:Pdelta}) should work  well for all sky 
coverage, even though the assumption of Gaussian $P_\delta$ breaks 
in the small sky coverage limit.  
The Poisson limit requires 
\ba
{\rm Re}(G)f_{\rm sky}\simeq {\rm Re}(\sum_j D_j)\, f_{\rm sky}\ll 1 \
    . \nonumber 
\ea
Fig. \ref{fig:check} shows that only for $f_{\rm sky}\ll 10^{-5}$,  
or the sky coverage $\ll 0.4$ deg$^2$, will the Poisson limit be reached.  
In reality, the above Poisson limit expression of $P(C)$ has very 
limited application, because all ongoing and upcoming SZ surveys have  $f_{\rm
    sky}\gg 10^{-5}$.

\end{document}